\documentclass[preprintnumbers,amsmath,amssymb]{revtex4}
\usepackage{graphicx}
\usepackage{bm}

\begin{document}

\title{
Characteristics of ion beams by laser acceleration in the Coulomb explosion regime
}

\author{Toshimasa Morita}

\affiliation{
Quantum Beam Science Research Directorate,
National Institutes for Quantum Science and Technology,
8-1-7 Umemidai, Kizugawa-shi, Kyoto 619-0215, Japan
}


\begin{abstract}
In the Coulomb explosion acceleration regime, an ion bunch with a narrow energy range
exhibits a thin shell shape with a certain diameter.
The ion cloud has a layered structure of these ion bunches with different energies.
The divergences of the ion bunch in the laser inclination direction and
perpendicular to it are different in an oblique incidence laser.
This indicates that the smaller the spot diameter of the laser pulse,
the larger the divergence of the ion beam.
In addition, theoretical formulas for the radius of the generated ion beam and
the energy spectrum are derived, and they are shown to be in good agreement with
the simulation results.

\end{abstract}


\maketitle

\section{Introduction}

The recent progress in compact laser systems has been significant.
Laser ion acceleration is a compelling application of high-power compact
lasers \cite{BWE,DNP}.
If a compact laser system can generate ions with sufficiently high energy and quality,
low-cost compact accelerators would become feasible.
However, the ion energies that have been currently achieved by laser ion acceleration
are insufficient for applications such as particle therapy \cite{WDB,HGK}.
Therefore, there are two ways for the study of the laser ion acceleration to proceed:
(i) evaluating the conditions required to produce higher energy ions
(including high power laser development)
\cite{BWP,DL,PRK,SPJ,Toncian,KKQ,MEBKY,TM1,TM2,Kiri}
and (ii) an investigation aimed at practical application using the currently available
ion energies.

In the former, the focus is on the maximum energy of the obtained ions.
In general, the higher the energy of ions, the smaller their number.
Therefore, the number of ions in an ion bunch, with a narrow energy width,
containing the maximum energy ion is very small.
However, numerous applications require an ion beam with a sufficient number of
ions \cite{ESI,BEE}.
Therefore, in this way,
it is necessary to satisfy the two conflicting requirements.

In this study,
we take the latter way, and it is assumed that a reasonable laser is used.
We focused not only on the maximum energy of the generated ions but
also on the lower energy ions, because they are produced in large quantities.
Our aim is to find a reasonable method to generate a large number of
not-so-high-energy ions (several MeV/u) within a narrow solid angle and to clarify
the characteristics of these ion beams. Such an ion beam can be applied, e.g.
as an injector in heavy-ion radiotherapy facilities.

A foil is used as the target because of its ease of preparation and handling.
Currently, carbon ions are most commonly used in heavy-ion radiotherapy.
Therefore, a carbon foil target is selected for this study.
Reasonable laser conditions, i.e. a relatively low laser energy and
sufficiently small spot size and pulse duration, are considered.
This is because the laser used in a compact accelerator will inevitably have
reasonable performance due to the requirements for a smaller size and
lower price of the accelerator.

The remainder of this paper is organized as follows:
In Section  \ref{para}, the simulation parameters are presented.
Section \ref{resul} presents the characteristics of the accelerated carbon ions.
The analytical considerations are presented in Section \ref{theory}.
Section \ref{resu_4m} presents the properties of the generated carbon ion beams at
an energy of around $4$ MeV/u.
Section \ref{con} summarizes the main results of this study.

\section{Simulation model} \label{para}

Simulations were performed using a parallelized electromagnetic code based on
the particle-in-cell (PIC) method \cite{CBL}.
The parameters used in the simulations are as follows.

An idealized model is used in which a Gaussian $p$-polarized laser pulse is
obliquely incident, $45^{\mathrm{o}}$, on a foil target represented by
a collisionless plasma.
The laser pulse energy is  $2.5$ J and focused to a spot size of $5 \mu$m full width
at half maximum (FWHM), and the pulse duration is $50$ fs (FWHM).
Corresponding to the laser pulse with
peak intensity, $I$, is $1.7\times 10^{20}$ W/cm$^{2}$, the peak power is $47$ TW,
and dimensionless amplitude $a_0=q_eE/m_{e}\omega c=9$.
The laser wavelength is $\lambda = 0.8 \mu $m.

A foil target consisting of carbon is used in this study.
It has been reported that high energy ions are generated at $0.1 \mu$m foil thickness
when an $18$ J laser pulse is irradiated onto
a polyethylene (CH$_2$) foil target \cite{TM2}.
Because the energy of the laser pulse used in this study is much lower than that of
the aforementioned laser pulse, and moreover, the electron density in the fully
striped carbon target is about two times that of CH$_2$,
the foil thickness at which the maximum energy occurs is considered to be much thinner
than $0.1 \mu$m.
However, it is difficult to achieve extremely thin thicknesses.  Therefore, the foil
thickness, $\ell_t$, is set to $1 \mu$m as the thinnest thickness considered feasible.

The ionization state of the carbon ion is assumed to be $Z_{i}=+6$.
The electron density is $n_{e}=6\times 10^{23}$ cm$^{-3}$, i.e.
$n_{e}=345 n_\mathrm{cr}$, where $n_\mathrm{cr}$ denotes the critical density.
The total number of quasiparticles is $2 \times 10^{11}$.

The number of grid cells is $4000 \times 3456 \times 2500$ along
the $X$, $Y$, and $Z$ axes.
Correspondingly, the simulation box size is $72\mu$m$ \times 62\mu$m$ \times 45\mu$m.
The boundary conditions for the particles and fields are periodic in the
transverse ($Y$, $Z$) directions and
absorbing at the boundaries of the computation box along the $X$ axis.
We set the electric field of the laser pulse to be along the $XY-$plane,
so that the magnetic field of the laser pulse is along the $Z$ axis.
The laser-irradiated side of the foil surface is placed at $X=36 \mu$m,
and the center of the laser pulse is located $13 \mu$m in front of its surface
in the $X$ direction and $13 \mu$m above the foil center in the $Y$ direction.
The laser propagation direction is $45^{\mathrm{o}}$ downward,
i.e. the direction vector is $(1,-1,0)$.
An $xyz-$coordinates system is used throughout the text and figures.
The origin of the coordinate system is located at the center of the laser-irradiated
surface of the initial target, and the directions of the $x$, $y$, and $z$ axes are
the same as those of the $X$, $Y$, and $Z$ axes, respectively.
Therefore, the $x$ axis denotes the direction perpendicular to the target surface,
and the $y$ and $z$ axes lie in the target surface.

\section{Accelerated carbon ions} \label{resul}

\begin{figure}[tbp]
\includegraphics[clip,width=9.5cm,bb=27 27 460 424]{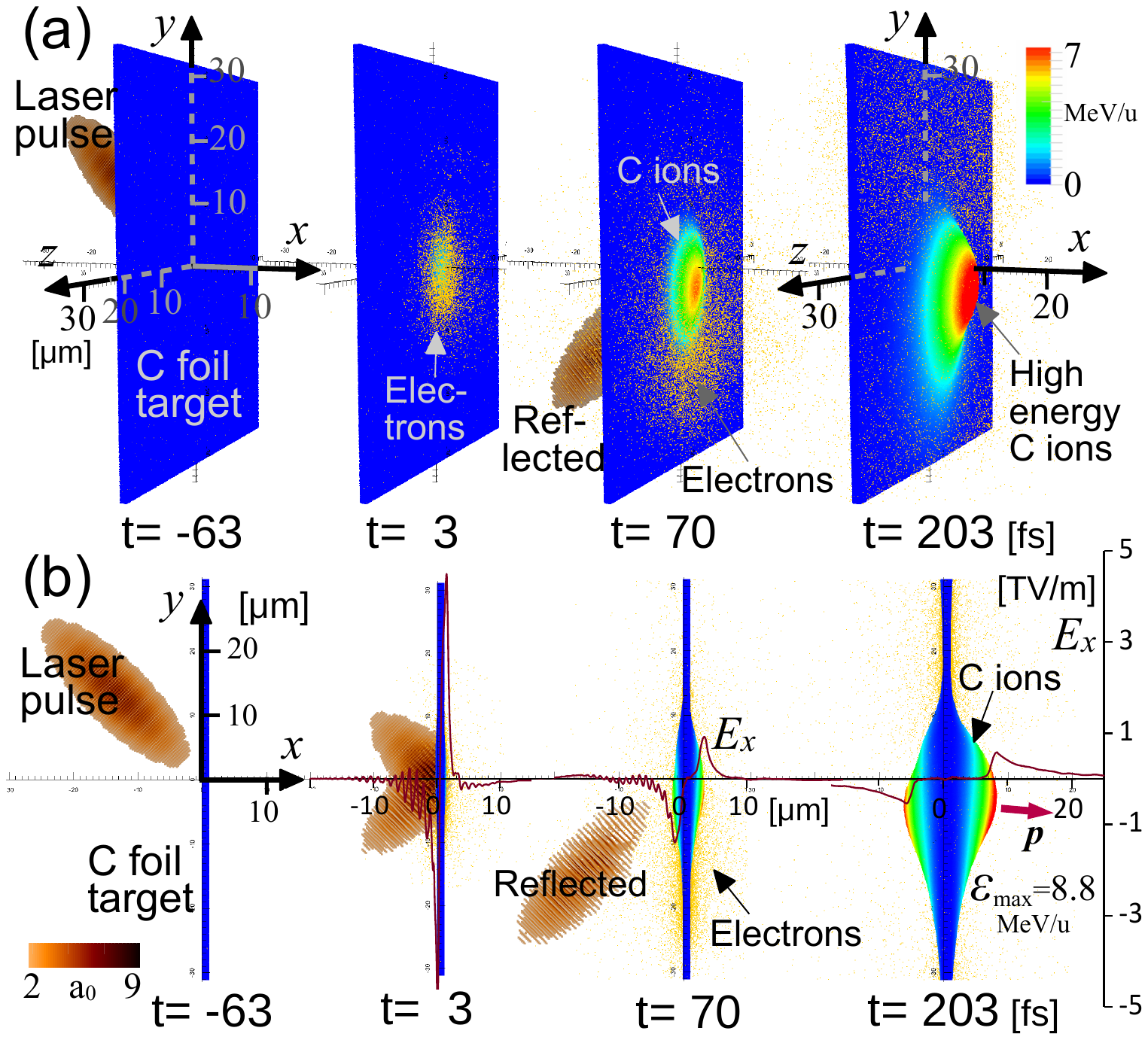}
\caption{
(a) Laser pulse is oblique incident, $45 ^{ \mathrm{o}}$, on carbon foil target.
A 3D view is given of the particle distribution and electric field magnitude
(isosurface for value $a_0=2$) at each time.
Half of the electric field box has been excluded to reveal the internal structure.
The carbon ions color corresponds to their energy.
(b) A 2D projection is shown as viewed along the $z$ axis.
Here, half of the ion cloud also has been excluded.
The red thick arrow shows the momentum vector, $\bm{p}$,
of the high energy carbon ions.
The $x$-component of the electric field, $E_x(x)$, on the $x$ axis are
shown by solid lines.
}
\label{fig:fig_pdis}
\end{figure}

In this section, we present the simulation results.
Under our conditions, because $I/n_e \ell_t = 3$ W/electron,
which is considerably less than $I/n_e \ell_t > 500$ W/electron
which is a condition that would be non-homeomorphic acceleration \cite{TM3},
therefore the ion acceleration scheme is the Coulomb explosion.
Figure \ref{fig:fig_pdis} shows the particle distribution and electric field
magnitude at times $t= -63, 3, 70$, and $203$ fs.
Half of the electric field box has been excluded to reveal the internal structure
in Fig. \ref{fig:fig_pdis}(a), and
their cross sections in the $xy-$plane are shown in Fig. \ref{fig:fig_pdis}(b).
Carbon ions are classified by color in terms of their energy.
It is assumed that $t=0$ when the center of the laser pulse, where $I$ is
the strongest, reaches the laser-irradiated surface of the initial target.
That is,
more than half of the laser pulse does not interact with the target when $t<0$,
and more than half of the interaction is complete at $t \geq 0$.
The simulation start time is $t=-63$ fs.
The initial shapes of the laser pulse and target are shown at $t=-63$ fs.
The laser pulse is defined on the $-x$ side of the target and
travels diagonally toward the $+x$ side.
At $t=3$ fs, the laser pulse undergoes strong interactions with the target.
About half of the laser pulse has interacted with the target, whereas
the other half does not.
A portion of the laser pulse is reflected from the target.
Subsequently, the carbon ion cloud explodes by Coulomb explosion and grows over time.
High energy carbon ions are distributed at the $+x$ side tip of the ion cloud and
shifted slightly in the $-y$ direction.

At $t=70$ fs, the interaction between the laser pulse and target ended,
and the reflected laser pulse is moving in the $-x$ direction (diagonally downward).
The target is slightly expanded. In the electron distribution,
the electrons that are pushed out from the target are distributed in large numbers in
the $-y$ region \cite{MEBKY2}.
At $t=203$ fs, the target is largely expanded, with high energy areas occurring
at each tip of the ion cloud on the $+x$ and $-x$ sides.
The maximum energy of the carbon ions is $8.8$ MeV/u and
occurs at the tip of the $+x$ side of the ion cloud.
The dark-red arrow represents the momentum vector of the high energy carbon ions.
This momentum vector is slightly tilted downward,
in the $-y$ direction \cite{MEBKY,MEBKY2},
and its angle with the $x$ axis is $-5^{\mathrm{o}}$.
Although it is lower than that of the $+x$ side, relatively high energy ions
are also generated at the tip of the ion cloud on the $-x$ side,
with a maximum energy of $7.1$ MeV/u.
These ions travel in the opposite direction to the ions on the $+x$ side,
i.e. in the $-x$ direction.
High energy ions also appear on the $-x$ side because the acceleration scheme in
this study is the Coulomb explosion of the target.
The distribution of $E_x$ on the $x$ axis is shown as a solid line.
The ions are accelerated by $E_x$.
The largest $E_x$ in the figures occurs on the $+x$ side surface of the target
at $t=3$ fs, which is $5$ TV/m.
At this time, a similarly large $E_x$ also occurs on the $-x$ side surface in the
opposite direction.
Therefore, it can be said that ion acceleration mainly occurs at around $t=3$ fs when
the laser and target interact strongly.
Subsequently, $E_x$ is remarkably small and decreases over time.
At $t=203$ fs,
it is observed that the $E_x$ inside the ion cloud is approximately $0$.

\begin{figure}[tbp]
\includegraphics[clip,width=9.0cm,bb=50 54 552 275]{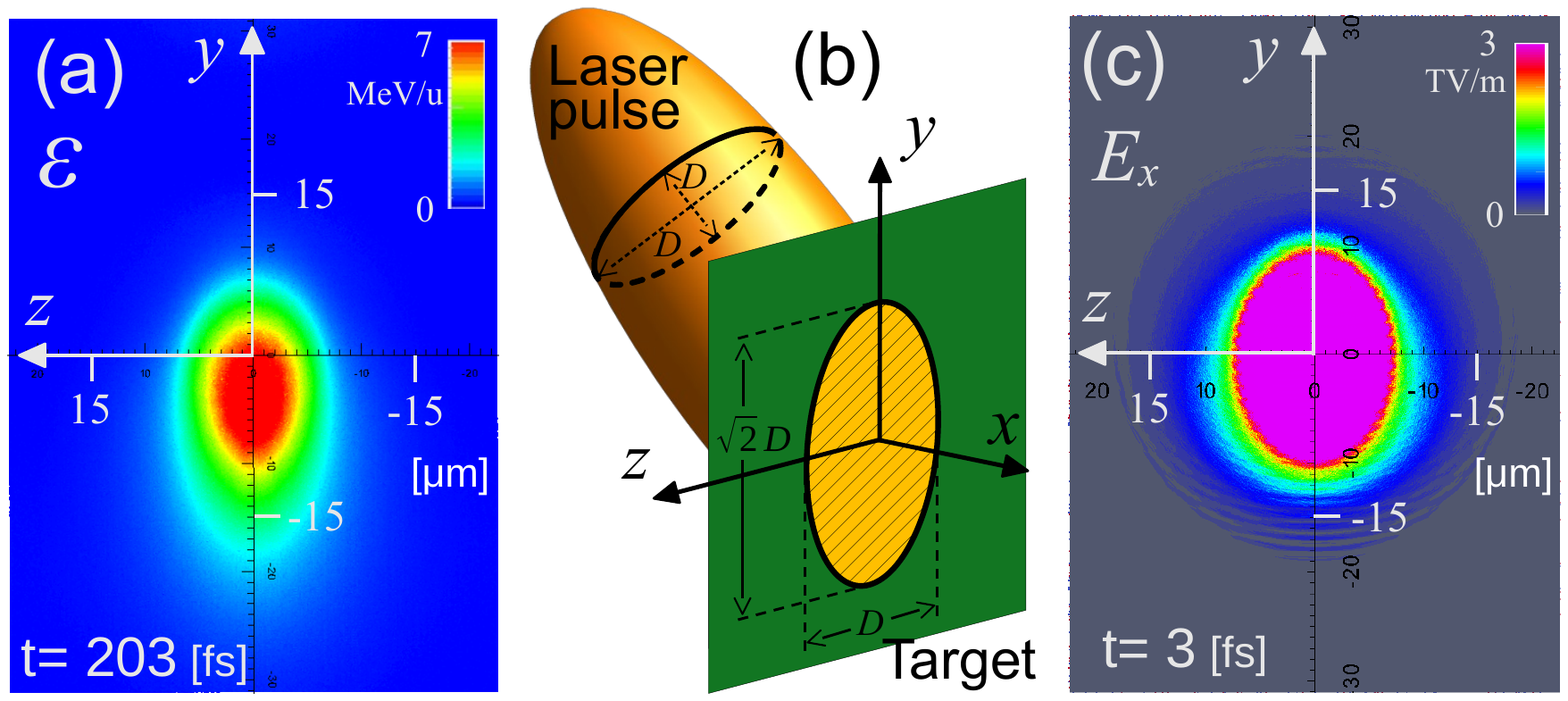}
\caption{
(a) Carbon ion distribution at $t=203$ fs.
A 2D projection of Fig. \ref{fig:fig_pdis}(a) is shown as viewed along the $x$ axis.
The high energy ions are distributed in a vertically long elliptical shape.
(b) Illustration of the laser pulse and target.
The cross-section of the laser pulse is a circle,
but the distribution of the laser on the target is a vertically long ellipse.
(c) $E_x$ distribution on the target at $t=3$ fs.
$E_x$ has a vertically long elliptical distribution.
The purple area is more than $3$ TV/m.
}
\label{fig:fig_lsr}
\end{figure}

In the following, we consider the characteristics of the generated ion cloud.
First, the spatial distribution of the ion cloud is considered.
Figure \ref{fig:fig_lsr}(a) shows a two dimensional (2D) view of the ion distribution
at $t=203$ fs in Fig. \ref{fig:fig_pdis}(a), as viewed along the $x$ axis.
The high energy ions are distributed in a long elliptical shape in the $y$ direction,
and its center is shifted toward the $-y$ direction.

Hereinafter, we refer to the $y$ axis direction as the vertical and
the $z$ axis direction as the horizontal.
The generated ion distribution is a vertically long ellipse because the laser is
an oblique incidence inclined in the $y$ direction.
In our simulation, the cross-section of the laser pulse is circular,
but its shape on the target surface is elliptical because of oblique incidence
(see Fig. \ref{fig:fig_lsr}(b)).
Consequently, the distribution of the acceleration field generated on the target
also becomes elliptical (Fig. \ref{fig:fig_lsr}(c)). 
Figure \ref{fig:fig_lsr}(c) shows the distribution of the $x$ direction electric field
(accelerating field), $E_x$, on the $+x$ side surface of the target at $t=3$ fs,
which is the time when the strongest $E_x$ occurs.
Because ions are produced from this elliptically distributed acceleration field,
their distribution becomes elliptical.
From the details shown in Fig. \ref{fig:fig_lsr}(c),
it can be seen that the accelerating electric field is egg-shaped.
That is, the left and right sides are symmetrical with respect to the $y$ axis,
but the top and bottom sides are not symmetrical with respect to the $z$ axis, and
they are distributed horizontally longer in the region where $y<0$ than
in the region where $y>0$.
In addition, from the $y$ coordinate values of the light blue area,
it is clear that this area is distributed farther in the $-y$ direction than
in the $+y$ direction.
This is because the laser is obliquely incident; thus, more electrons are distributed
in the area of $y<0$ outside the target. (see $t=70$ fs in Fig. \ref{fig:fig_pdis}).
The accelerated ions generated from this elliptical region at $t=3$ fs are 
thereafter more shifted toward the $-y$ direction due to the effect of the electron
distribution \cite{MEBKY2}.

\begin{figure}[tbp]
\includegraphics[clip,width=8.0cm,bb=29 28 367 238]{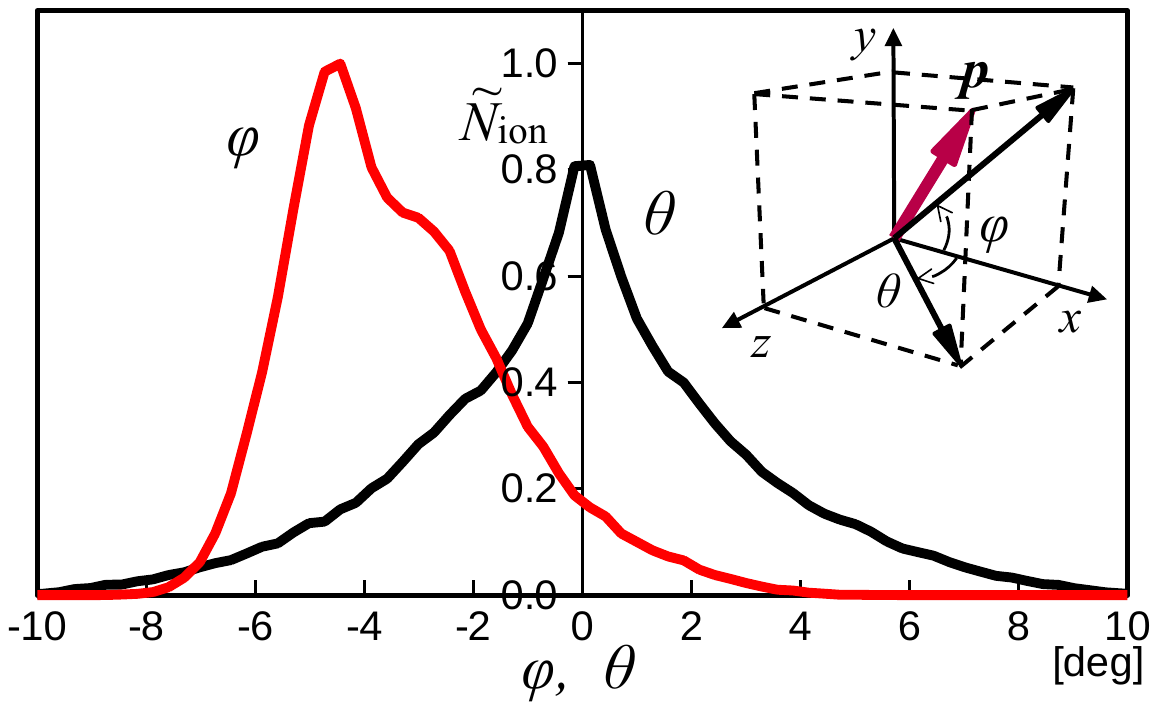}
\caption{
Frequency distribution of $\varphi$ and $\theta$ of the ions of $3.8$ MeV/u or
more traveling in the $+x$ direction at $t=203$ fs.
It is normalized by the maximum value of $\varphi$,
$\tilde{N}_\mathrm{ion}=N_\mathrm{ion}/N_\mathrm{ion,max}$.
The momentum vector of an ion is tilted downward ($-y$ direction) on average,
but not horizontally.
The horizontal spread of the momentum vector is wider than the vertical direction.
In the inset: the momentum vector of an ion, $\varphi$, and $\theta$.
}
\label{fig:fig_theta}
\end{figure}

Next, the traveling direction of the generated ions, i.e. the direction of
their momentum vectors, is considered.
Figure \ref{fig:fig_theta} shows the frequency distribution of the angle between
the momentum vector of an ion of $3.8$ MeV/u or more that is accelerating
toward $+x$ side and the $x$ axis at $t=203$ fs.
Energies of $3.8$ MeV/u or more are selected because we are focusing on carbon ions
at around $4$ MeV/u (to be discussed in detail in Section \ref{resu_4m}).
$\varphi$ is the angle between the momentum vector projected onto the $xy-$plane and
the $x$ axis, i.e. the vertical tilt, and $\theta$ is the angle when projected onto
the $xz-$plane, i.e. the horizontal tilt (see the inset).
Modes are $\varphi_\mathrm{mode}$=$-4.4^{\mathrm{o}}$ and
$\theta_\mathrm{mode}$=$0^{\mathrm{o}}$,
that is, the traveling directions of these ions are, on average, slightly tilted
in the $-y$ direction, vertical direction, but not in the $z$ direction,
horizontal direction.
The vertical direction, $\varphi$, is distributed in $-8$ -- $4^{\mathrm{o}}$,
and the horizontal direction, $\theta$, is distributed in approximately
$-10$ -- $10^{\mathrm{o}}$.
Their widths are $\Delta\varphi$ = $12^{\mathrm{o}}$ and
$\Delta\theta$ = $20^{\mathrm{o}}$, respectively, i.e.
the horizontal spread of the velocity vector is wider than the vertical direction.
That is, the ion cloud is expanding more in the horizontal direction.

\begin{figure}[tbp]
\includegraphics[clip,width=8.0cm,bb=37 33 543 360]{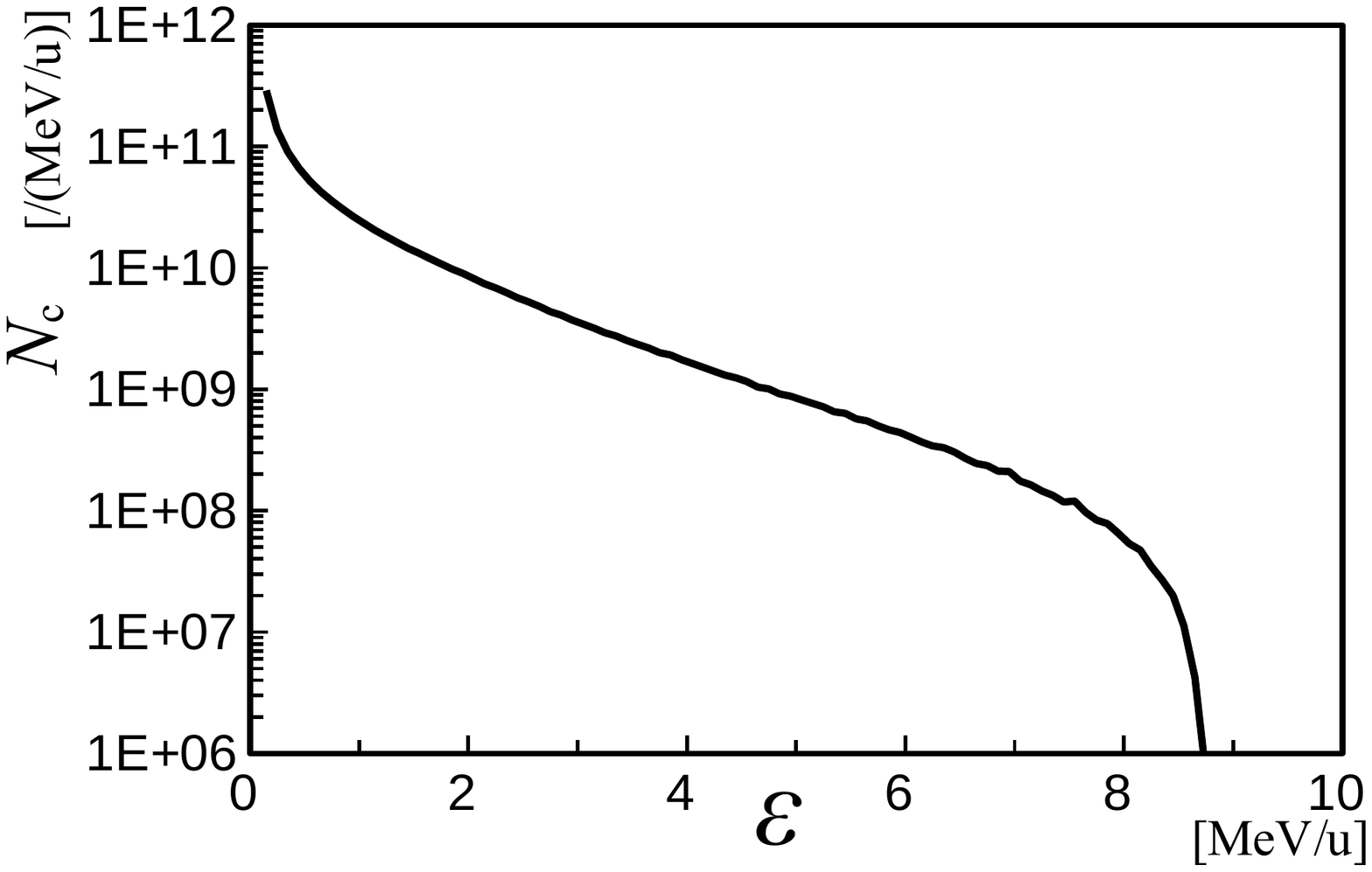}
\caption{
Energy spectrum of the carbon ions traveling in the $+x$ direction obtained
in the simulation at $t=203$ fs.
It is a monotonically decreasing.
}
\label{fig:fig_espc}
\end{figure}

Figure \ref{fig:fig_espc} shows the energy spectrum of carbon ions accelerated in
the $+x$ direction at $t=203$ fs.
Carbon ions are generated on both the $+x$ and $-x$ sides of the target, but here,
it is assumed that the ions are received and used by some device installed behind
the target, i.e. in the $+x$ direction.
Therefore, only the ions that traveled to the $+x$ side are considered.
The vertical axis is provided in units of the number of carbon ions per
$1$ MeV/u width.
The horizontal axis represents the energy of the carbon ion $\mathcal{E}$.
The number of ions produced decreases with increasing energy.
Therefore, if a large number of accelerated ions are needed,
ions in the lower energy range should be used, not near the maximum energy.
This is the reason for our focus on ions in the low energy region.
At $8$ MeV/u, which is near the maximum energy, there are $6 \times 10^7$ /(MeV/u)
ions, but at $4$ MeV/u, which is half the maximum energy,
there are $2 \times 10^9$ /(MeV/u) ions, which is about $30$ times the number of ions.
The number of ions with a $10\%$ energy range ($3.8$ -- $4.2$ MeV/u) at $4$ MeV/u is
approximately $7 \times 10^8$, which is sufficient for some applications
(e.g. particle therapy).
In Section \ref{resu_4m},
the ions generated in this energy range are discussed in detail.

In our simulation,
because the acceleration scheme is a Coulomb explosion due to the low laser intensity
and the intensity distribution is Gaussian,
the energy spectrum monotonically decreases.
The next section discusses this issue in more detail.

\section{Analytical consideration} \label{theory}

We consider the Coulomb explosion regime, i.e. TNSA \cite{WLC},
as the acceleration scheme,
similar to that shown above.
In this case, the maximum ion energy is generated from the opposite side surface
of the laser irradiation surface of the target, and from there,
the generated ion energy decreases with depth toward the inside of the target,
and near the center of the target thickness,
the ion energy is approximately zero.
In the region from there to the laser-irradiated surface,
the ion is accelerated in the opposite direction, $-x$ direction,
and the ion energy increases again to the laser-irradiated surface.

In this study, we focus on ions accelerated to the $+x$ side.
Therefore, we consider the area from near the center of the target thickness
to the target surface on the opposite side of the laser-irradiated surface.
In this section, this surface will be simply described as the target surface.

\subsection{Ion bunch radius} \label{theo_1}

The intensity, $I$, distribution of the laser pulse is Gaussian in the direction
 perpendicular to the laser traveling direction in our simulation.
Therefore, when the origin is at the position on the target surface where
the laser center is located, the laser electric field in any direction on
the target surface from the origin is a Gaussian distribution.
That is, at a certain time, the laser electric field at the $r$ position on
the surface of the target is given by
\begin{equation}
E_l(r)=E_l(0)e^{-ar^2},
\label{e_las}
\end{equation}
where $E_l$ is the amplitude of the electric field, $|\bm{E}_l|$, $E_l(0)$ is
its value at the origin, i.e. at the laser pulse center, and '$a$' is the coefficient
that defines the shape of the Gaussian distribution,
which is determined from the distribution shape of the laser intensity, $I$.

It is considered that the larger the laser electric field, the larger
the accelerating field generated on the target by laser irradiation.
Then, we assume that the accelerating electric field $E_s(r)$ generated at position
$r$ on the target surface by laser irradiation in the direction perpendicular to
the surface is proportional to the laser electric field $E_l(r)$ at that position.
That is, it is assumed that
$E_l(r)=k_e E_s(r)$,
where $k_e$ is a proportional constant.
By substituting this into  Eq. (\ref{e_las}), we obtain the following relational
equation for the accelerating electric field on the target surface:
\begin{equation}
E_s(r)=E_s(0)e^{-ar^2}.
\label{e_tar}
\end{equation}

Let $\mathcal{E}(r)$ be the maximum ion energy generated from the $r$ position and
$I(r)$ be the laser intensity at that position.
It was shown that when the accelerating process is a Coulomb explosion of the target,
i.e. TNSA, the maximum ion energy, $\mathcal{E}_\mathrm{max}$, is proportional
to the square root of the maximum laser intensity,
$\sqrt{I_\mathrm{max}}$ \cite{TM3}.
Here, $\mathcal{E}_\mathrm{max}$ is generated from the origin,
i.e. $\mathcal{E}_\mathrm{max}=\mathcal{E}(0)$, and $\sqrt{I_\mathrm{max}}$ is also
the value at the origin, i.e. $\sqrt{I_\mathrm{max}}=\sqrt{I(0)}$.
That is, there is an $\mathcal{E}(0) \propto \sqrt{I(0)}$ relationship.
Generalizing this, we assume here that the maximum ion energy, $\mathcal{E}(r)$,
generated from position $r$ is proportional to the square of the laser intensity
at that position, $\sqrt{I(r)}$, i.e. $\mathcal{E}(r) \propto \sqrt{I(r)}$.
Then, from the relation $\sqrt{I(r)} \propto E_l(r) \propto E_s(r)$, we obtain
\begin{equation}
E_s(r)=k_\epsilon\mathcal{E}(r),
\label{e_ene}
\end{equation}
where $k_\epsilon$ is a proportionality constant.
By substituting this into Eq. (\ref{e_tar}), we obtain
$\mathcal{E}(r)=\mathcal{E}(0)e^{-ar^2}$.
Solving this with $r$ and rewriting $\mathcal{E}(0)$ as $\mathcal{E}_\mathrm{max}$,
we obtain
\begin{equation}
r=\sqrt{-\frac{1}{a}\log\frac{\mathcal{E}(r)}{\mathcal{E}_\mathrm{max}}},
\label{r_ene}
\end{equation}
where $r$ indicates the distance from the origin to the position where the maximum
energy of the generated ion is $\mathcal{E}(r)$.
When a certain amount of energy, $\mathcal{E}$, is provided, $r$ is determined
by substituting $\mathcal{E}(r)$ with $\mathcal{E}$ in Eq. (\ref{r_ene}).
From the area within this $r$, ions with energy higher than $\mathcal{E}$ are produced
and $\mathcal{E}$ ions are also produced.
However, farther than $r$, no ions with energy greater than $\mathcal{E}$ occur.
Thus, $r$ is the maximum distance from the origin to the position where
ions of energy $\mathcal{E}$ occur.
In other words, the radius of the ion bunch with energy $\mathcal{E}$ is given by $r$.

\begin{figure}[tbp]
\includegraphics[clip,width=9.0cm,bb=9 64 570 359]{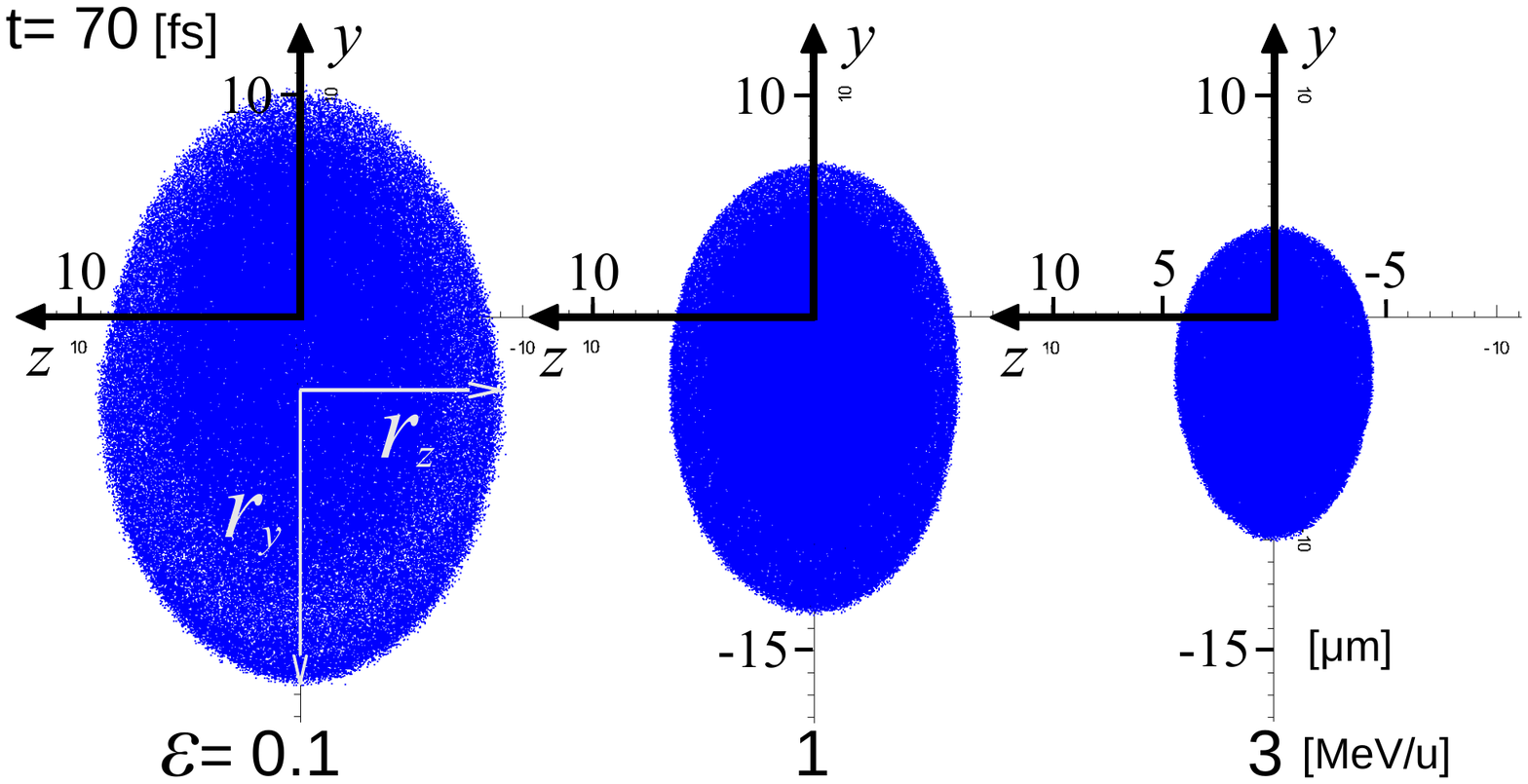}
\caption{
2D projections of ion bunches with energies of $0.1$, $1$, and $3$ MeV/u at $t=70$ fs
are shown as viewed along the $x$ axis.
The ion energy range for each bunch is $\pm5\%$.
Vertical radius, $r_y$, and horizontal radius, $r_z$, are shown.
Ion bunches have a vertically long shape.
The higher the energy, the smaller the radii.
}
\label{fig:fig_r_2d}
\end{figure}

Figure \ref{fig:fig_r_2d} shows a 2D view of the ion bunches with energies of
$0.1$, $1.0$, and $3.0$ MeV/u, as viewed along the $x$ axis, immediately after the
interaction between the laser pulse and the target ends ($t=70$ fs).
Here, the ion energy range is $\pm5\%$, that is, a $0.1$ MeV/u ion is an ion
with energy in the range of $0.095$ -- $0.105$ MeV/u and the same for other energies.
The distances from the center to the edge of the ion bunch in the $y$ and $z$
directions are defined as the vertical radius $=r_y$ and horizontal radius $=r_z$,
respectively (see Fig. \ref{fig:fig_r_2d}).
Because the laser is obliquely incident,
the ion bunch has a long vertical distribution, $r_y > r_z$.
The radii $r_y$ and $r_z$ are smaller for higher energy ion bunches.
The distribution of the $4$ MeV/u ion bunch will be shown in detail later
(in Sec. \ref{resu_4m}), and the ion bunches shown here are similar to that,
i.e. the thickness is thin and shaped like a shell.
The outlines of each ion bunch are clear and are not distributed in such a way that
there are no boundaries as the number of ions gradually becomes increasingly sparser.

\begin{figure}[tbp]
\includegraphics[clip,width=8.0cm,bb=33 38 312 296]{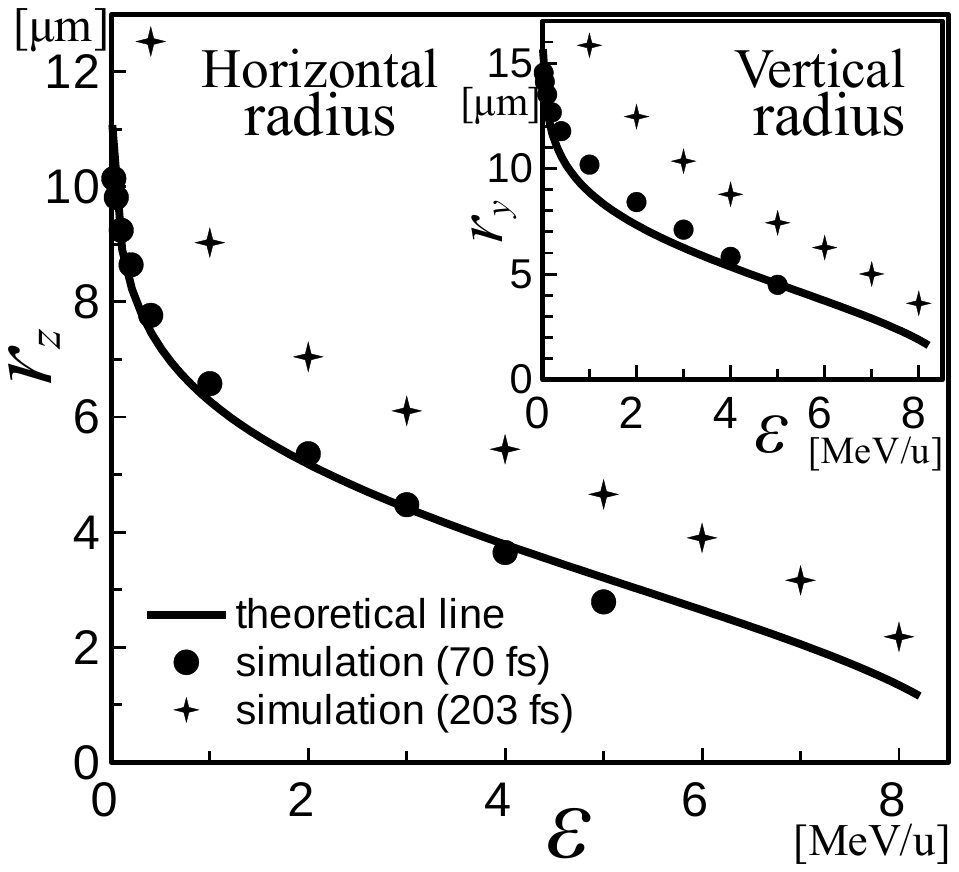}
\caption{
Simulation and theoretical results of horizontal radius, $r_z$, of carbon ion bunch
at each energy.
In the inset: their vertical radius, $r_y$.
}
\label{fig:fig_rene}
\end{figure}

Although the geometry of the ion bunches at only three energies has been shown,
more energies, $\mathcal{E}$, and radii, $r$, are shown in Fig. \ref{fig:fig_rene}.
The results of the simulation and theoretical formula in Eq. (\ref{r_ene}) are shown.
In this study, the Gaussian distribution shapes in the vertical and horizontal
directions are different, i.e. '$a$' is different,
because of the oblique incidence; thus results are shown for each direction.
The theoretical and simulation results of $r$ are in good agreement in both
the vertical and horizontal directions at $t=70$ fs,
immediately after the laser and target interaction ended.
This good agreement indicates that the two assumptions in the derivation of
the $r$ equation are correct.
$t=203$ fs, the Coulomb repulsion between ions in the ion cloud causes the ion bunch
to expand; consequently,
the simulation results are larger than the theoretical results.

In above consideration, we showed the case where the distribution of $I$ is Gaussian.
On generalizing Eq. (\ref{e_las}), when it is given by $E_l(r)=E_l(0)f(r)$,
using the function $f(r)$ for the distribution shape, Eq. (\ref{r_ene}) becomes
$r=f^{-1}(\mathcal{E}(r)/\mathcal{E}(0))$,
where $f(r)$ is assumed to be normalized to $f(0)=1$.

\subsection{Energy spectrum} \label{theo_2}

Next, a theoretical study on the number of generated ions is present.
For simplicity, we consider the normal incidence of the laser.

First, we determine the volume of the region in the initial target that will become
 the ion at $\mathcal{E}_1$ -- $\mathcal{E}_1{+\Delta\mathcal{E}}$ energy.
Let $r_1$ be the position on the target surface where an ion with
$\mathcal{E}_1$ energy appears from the surface.
That is, for $r > r_1$, $\mathcal{E}(r) < \mathcal{E}_1$ and
ions of $\mathcal{E}_1$ energy do not occur.

At a certain position $r$ on the target surface, where $r<r_1$, the thickness
direction, i.e. the depth direction, of the target is considered.
The origin is placed at the center of the target thickness,
and the axis is taken from this point toward the target surface.
We consider an electric field, i.e. an accelerating field,
oriented in the direction of this axis.
It is assumed that the electric field is zero at the origin,
and it at each position from there to the surface changes linearly to the surface
electric field, $E_s(r)$.
At $r$, the ions of $\mathcal{E}_1$ -- $\mathcal{E}_1{+\Delta\mathcal{E}}$ energy
are generated inside the target.
Let $E_1$ be the electric field at the position where the ions that become the
energy of $\mathcal{E}_1$ are located, $E_1{+\Delta E}$ be the electric field at the
position where the ions that become the energy of $\mathcal{E}_1{+\Delta\mathcal{E}}$
are located, and the distance between them be $\delta_1(r)$.
That is, we assume that
the ions of $\mathcal{E}_1$ -- $\mathcal{E}_1{+\Delta\mathcal{E}}$ occur from the
position where the electric field is $E_1$ -- $E_1{+\Delta E}$.
At this time, there exist a relationship $E_s(r)/\ell=\Delta E/\delta_1(r)$.
Therefore,
\begin{equation}
\delta_1(r) = \frac{\ell \Delta E}{E_s(r)},
\label{e_delt}
\end{equation}
where $\ell$ is the distance from the origin to the target surface; here,
half of the target thickness, $\ell_t/2$, is considered.

Next, let the position on the target surface where the laser center is located be
the origin, and consider polar coordinates on the surface, let $r$ be the distance
from the origin and $\theta$ be the angle from the horizontal axis.
At position ($r$, $\theta$), the small area $\Delta A=r\Delta\theta \Delta r$ where
the bound is the small length $\Delta r$ and small angle $\Delta\theta$.
At this position, thickness $\delta_1(r)$, which becomes the energy of
$\mathcal{E}_1$ -- $\mathcal{E}_1{+\Delta\mathcal{E}}$, is determined by
Eq. (\ref{e_delt});
thus, its volume is
$\Delta V = \Delta A \delta_1(r)=r\Delta\theta \Delta r \ell \Delta E/E_s(r)$.
Consequently, the total volume, $V$, of the region that becomes the energy of
$\mathcal{E}_1$ -- $\mathcal{E}_1{+\Delta\mathcal{E}}$, is obtained by integrating
$\Delta V$ with the target region, $V_1$, in $r \leq r_1$, i.e.
integrating $\theta$ with $0$ -- $2\pi$ and $r$ with $0$ -- $r_1$,
\begin{equation}
V = \int_{V_1} \Delta V =
\int_0^{r_1} \int_0^{2\pi}\frac{\ell \Delta E}{E_s(r)}rd\theta dr \nonumber \\
 = 2\pi \ell \Delta E \int_0^{r_1}\frac{r}{E_s(r)}dr.
\label{vol}
\end{equation}
Substituting Eq. (\ref{e_tar}) into the above equation, we obtain
\begin{equation}
V = \frac{2\pi \ell \Delta E}{E_s(0)} \int_0^{r_1}re^{ar^2}dr
 = \frac{\pi \ell \Delta E}{aE_s(0)}\left(e^{a{r_1}^2} -1 \right) \nonumber \\
 = \frac{\pi \ell \Delta E}{a}\left(\frac{1}{E_s(r_1)}-\frac{1}{E_s(0)} \right).
\label{vol-2}
\end{equation}
Here, the relationship $e^{a{r_1}^2}=E_s(0)/E_s(r_1)$ obtained
from Eq. (\ref{e_tar}) is used.
When the number density of the ions in the target is $n_0$, the number of ions
at energy $\mathcal{E}_1$ -- $\mathcal{E}_1{+\Delta\mathcal{E}}$ is $n_0V$.
This is expressed as $N(\mathcal{E}_1)\Delta \mathcal{E}$ using the number of ions
per unit energy range, $N(\mathcal{E})$.
Therefore, $N(\mathcal{E}_1)\Delta \mathcal{E}=n_0V$.
Substituting Eq. (\ref{vol-2}) for $V$ in this equation,
and thereafter substituting the relations
$\Delta E=k_\epsilon\Delta\mathcal{E}$, 
$E_s(r_1)=k_\epsilon\mathcal{E}(r_1)=k_\epsilon\mathcal{E}_1$, and
$E_s(0)=k_\epsilon\mathcal{E}(0)=k_\epsilon\mathcal{E}_\mathrm{max}$
obtained from Eq. (\ref{e_ene}), we obtain
\begin{equation}
N(\mathcal{E}) = \frac{n_0\pi \ell}{a}
  \left( \frac{1}{\mathcal{E}}-\frac{1}{\mathcal{E}_\mathrm{max}} \right),
\label{espe}
\end{equation}
for the energy spectrum.
Here, $\mathcal{E}_1$ is rewritten as $\mathcal{E}$ in general notation.

If the target thickness, $\ell_t$, is sufficiently thin relative to the laser
intensity, $\ell=\ell_t/2$ can be considered.
However, in our simulation, $\ell_t$ is thick with $1\mu$m; thus,
if $\ell=\ell_t/2$, the results would be large, because relatively high energy ions
would be produced even from near a depth of $0.5\mu$m from the surface.
That is, in a relatively thick target, the position where the electric field on the
target surface decreases to zero is not at the thickness center position of the
target, but closer to the surface, and $\ell$ needs to be its value.

\begin{figure}[tbp]
\includegraphics[clip,width=8.0cm,bb=38 32 530 350]{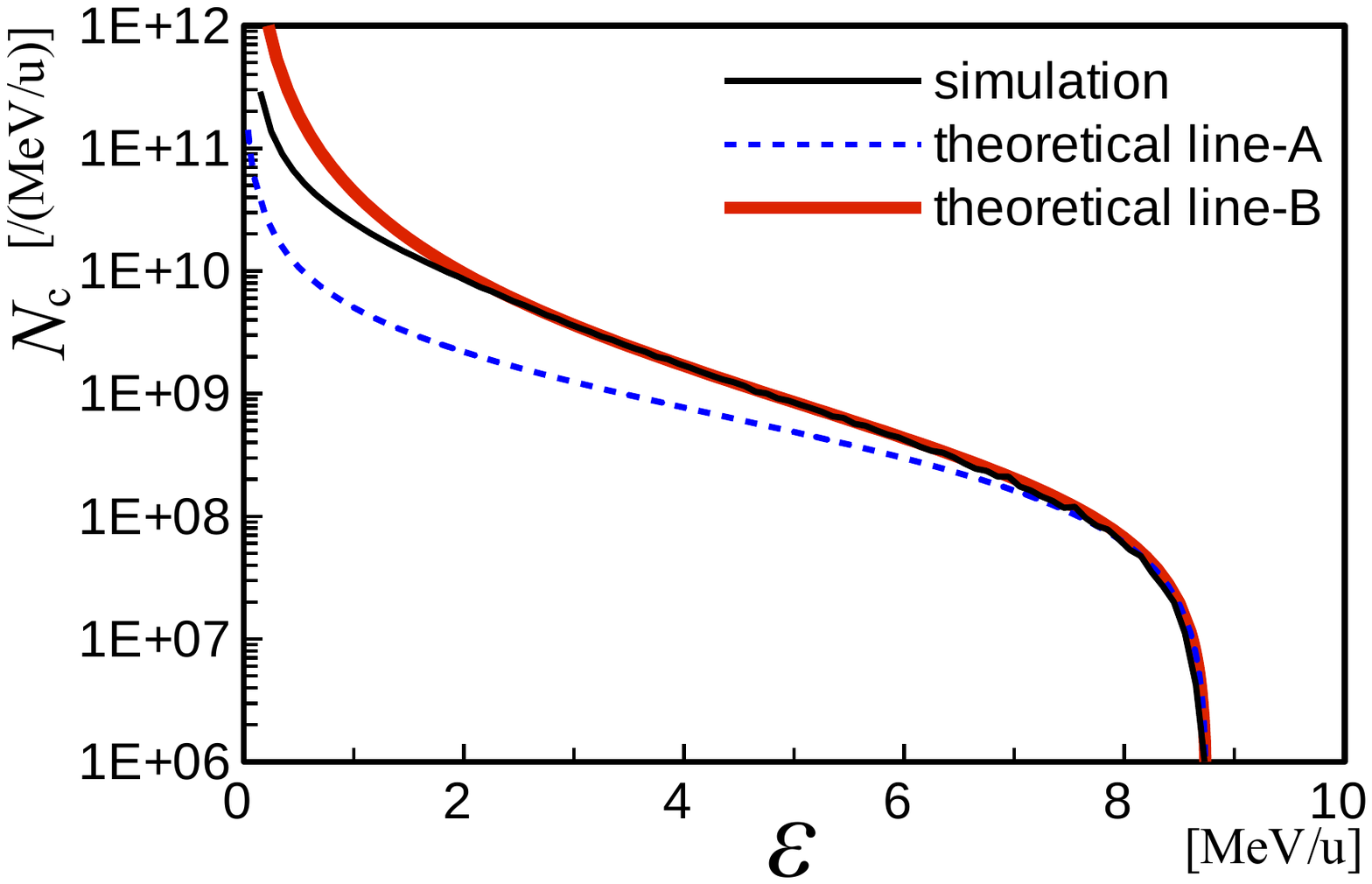}
\caption{
Energy spectrum in theory and simulation.
The dotted and red solid curves are given from
Eqs. (\ref{espe}) and (\ref{theo-2}), respectively.
}
\label{fig:fig_th_e}
\end{figure}

Figure \ref{fig:fig_th_e} line-A is the result for a value of $\ell$ of $1$nm.
That is, the part from the target surface to a depth of $1$nm will be accelerated.
The black solid line is the simulation result and is the same as shown
in Fig. \ref{fig:fig_espc}.
The theoretical and the simulation results agree well in the high energy range,
$\mathcal{E} \gtrsim 7$ MeV/u.

Rewriting Eq. (\ref{e_delt}) in terms of the ion energy using the relationship in
Eq. (\ref{e_ene}), we obtain $\delta_1(r) = \ell \Delta \mathcal{E}/\mathcal{E}(r)$.
This implies that regardless of the energy of interest,
if the energy range, $\Delta \mathcal{E}$, is the same,
the width of the region where ions of that energy occur is also the same.
This is because we assumed that the generated ion energy, the accelerating electric
field, in the target depth direction decreases linearly with depth.
In contrast, assume that the generating ion energy decreases exponentially
with $\xi$ (i.e. the accelerating electric field decreases exponentially) as
$\mathcal{E}(r)e^{-\xi\mathcal{E}(r)/\mathcal{E}_\mathrm{max}\ell}$,
where $\xi$ is the distance from the target surface; then, the region width is
$\delta_1(r)\mathcal{E}_\mathrm{max}/\mathcal{E}_1$,
and Eq. (\ref{espe}) is written as
\begin{equation}
N(\mathcal{E})=
\frac{n_0 \pi \ell}{a}
\frac{\mathcal{E}_\mathrm{max}-\mathcal{E}}{\mathcal{E}^2},
\label{theo-2}
\end{equation}
where $\ell$ is the distance from the surface to the position where the accelerating
electric field, the generated ion energy, of the surface is $1/e$ at $r=0$ which
is the center of the target surface.
Fig. \ref{fig:fig_th_e} line-B shows this theoretical result, and $\ell=1$nm,
as in line-A.
The theoretical and simulation results agree well, although
in the low energy range, $\mathcal{E} < 1.5$ MeV/u,
the theoretical results are slightly higher than the simulation results.

In Fig. \ref{fig:fig_th_e}, we set $\ell$ to $1$nm, but in other words,
from this result we can say that, under the conditions of our study,
those become carbon ions and accelerated by laser irradiation is at a depth of
approximately $1$nm from the target surface.
Eq. (\ref{e_delt}) indicates that the thickness, $\delta_1(r)$, at which the ions of
energy $\mathcal{E}_1$ are generated becomes thicker,
i.e. the number of generated ions increases, farther away from the origin.
This explains why the boundary of each ion bunch is clear,
as shown in Fig. \ref{fig:fig_r_2d}.

Differentiating Eq. (\ref{theo-2}) by $\mathcal{E}$, we obtain
$dN(\mathcal{E})/d\mathcal{E}=
n_0 \pi \ell(1-2\mathcal{E}_\mathrm{max}/\mathcal{E})/a\mathcal{E}^2 < 0$.
Therefore, the energy spectrum decreases monotonically.

\section{4MeV/u carbon ion bunch} \label{resu_4m}

\begin{figure}[tbp]
\includegraphics[clip,width=7.0cm,bb=30 32 344 459]{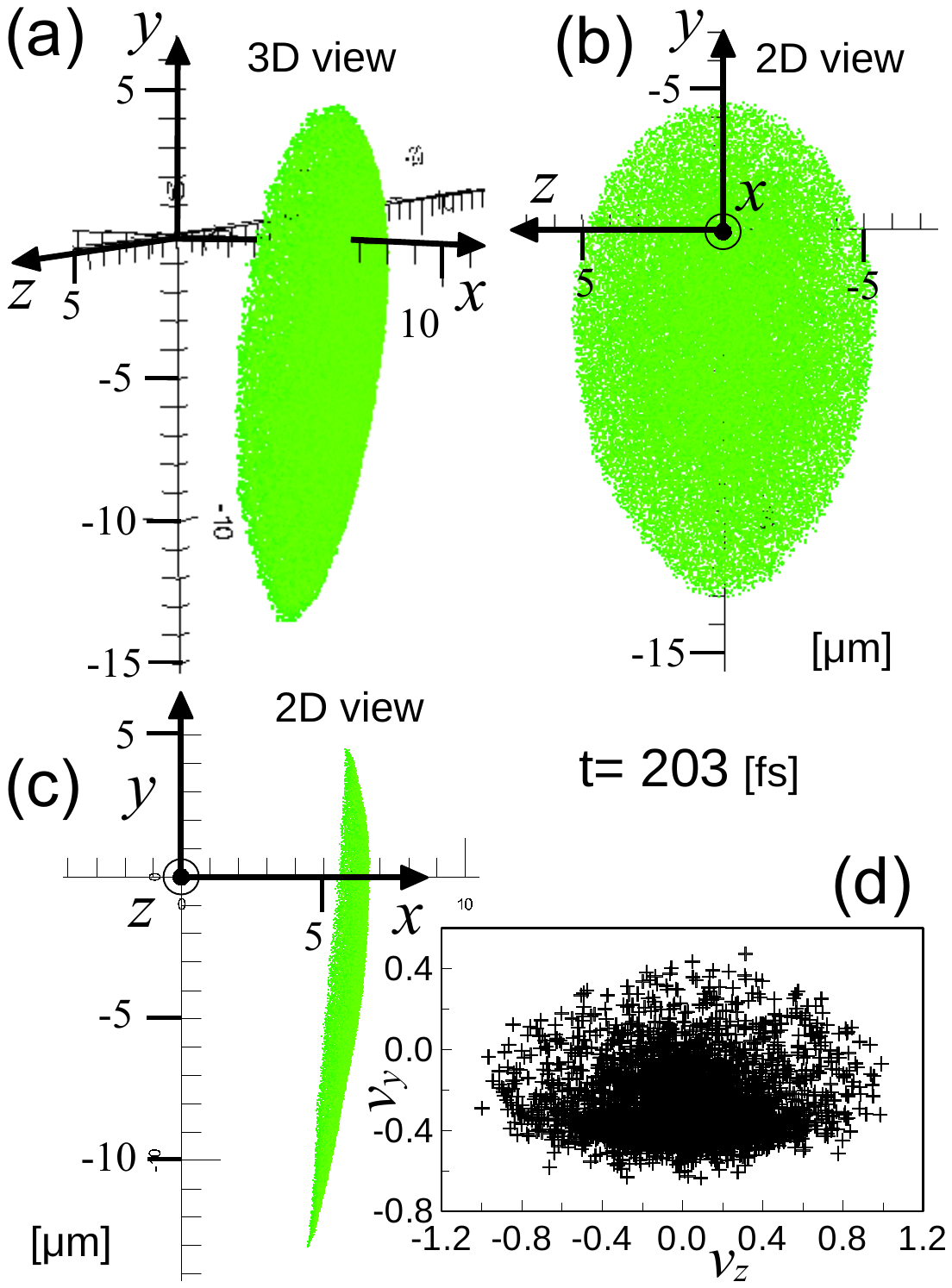}
\caption{
Spatial distributions of the $4$ MeV/u carbon ions at $t=203$ fs.
(a) 3D view.
2D views as viewed along the $x$ axis (b),
and viewed along the $z$ axis (c).
(d) Scatter plot of the $4$ MeV/u ion. $z$-direction velocity of carbon ions, $v_z$,
versus $y$-direction velocity, $v_y$, normalized by the maximum value of $|v_z|$.
}
\label{fig:fig_4mev}
\end{figure}

In this section,
we focus on ions around $4$ MeV/u, which is about half of the maximum ion energy,
because a large number of accelerated ions are obtained.
The energy range of the ion cloud shall be $10\%$ ($4$ MeV/u $\pm5\%$),
i.e. $3.8$ -- $4.2$ MeV/u ions.
In the following, unless otherwise noted, "$4$ MeV/u ion" means to a $3.8$ -- $4.2$
MeV/u carbon ion traveling in the $+x$ direction.
Figure \ref{fig:fig_4mev}(a,b,c) show the spatial distributions of the $4$ MeV/u
carbon ions at $t=203$ fs.
Figure \ref{fig:fig_4mev}(a) is a three dimensional (3D) view.
Figures \ref{fig:fig_4mev}(b) and (c) are 2D projections as viewed along the $x$ and
$z$ axes, respectively.
As shown in Fig. \ref{fig:fig_4mev}(b), the ion cloud has a vertical, $y$ direction,
long elliptical distribution, with a slight overall shift in the $-y$ direction.
The thickness of the ion cloud is very thin (see Fig. \ref{fig:fig_4mev} (c)) and
has a shell-like shape spread along the $yz-$plane.
This is because the acceleration process is a Coulomb explosion.
In the Coulomb explosion, the ion at the surface of the target becomes the maximum
energy ion, and the ion energy decreases as it moves inward from the surface in the
direction of the thickness (but only up to half the target thickness).
Therefore, an ion at a certain energy, $\mathcal{E}$, occurs only from
a certain position, $x$,
in the depth direction.
That is, the ions with energy of
$\mathcal{E}$ -- $\mathcal{E}{+\Delta\mathcal{E}}$
come from a narrow depth range of
$x$ -- $x+\Delta x$, so the thickness of its bunch becomes very thin.
The ion cloud has a layered structure consisting of shell-like thin ion bunches with
different energies, similar to the layers of an onion.
Figure \ref{fig:fig_4mev}(d) shows a scatter plot of the $4$ MeV/u ions when
the horizontal, $z$ direction, and vertical, $y$ direction, velocities are
taken on the horizontal and vertical axes, respectively.
The ions are long distributed in the horizontal direction, i.e.
the ion cloud has a larger horizontal velocity spread than in the vertical direction.

\begin{figure}[tbp]
\includegraphics[clip,width=9.0cm,bb=37 29 515 281]{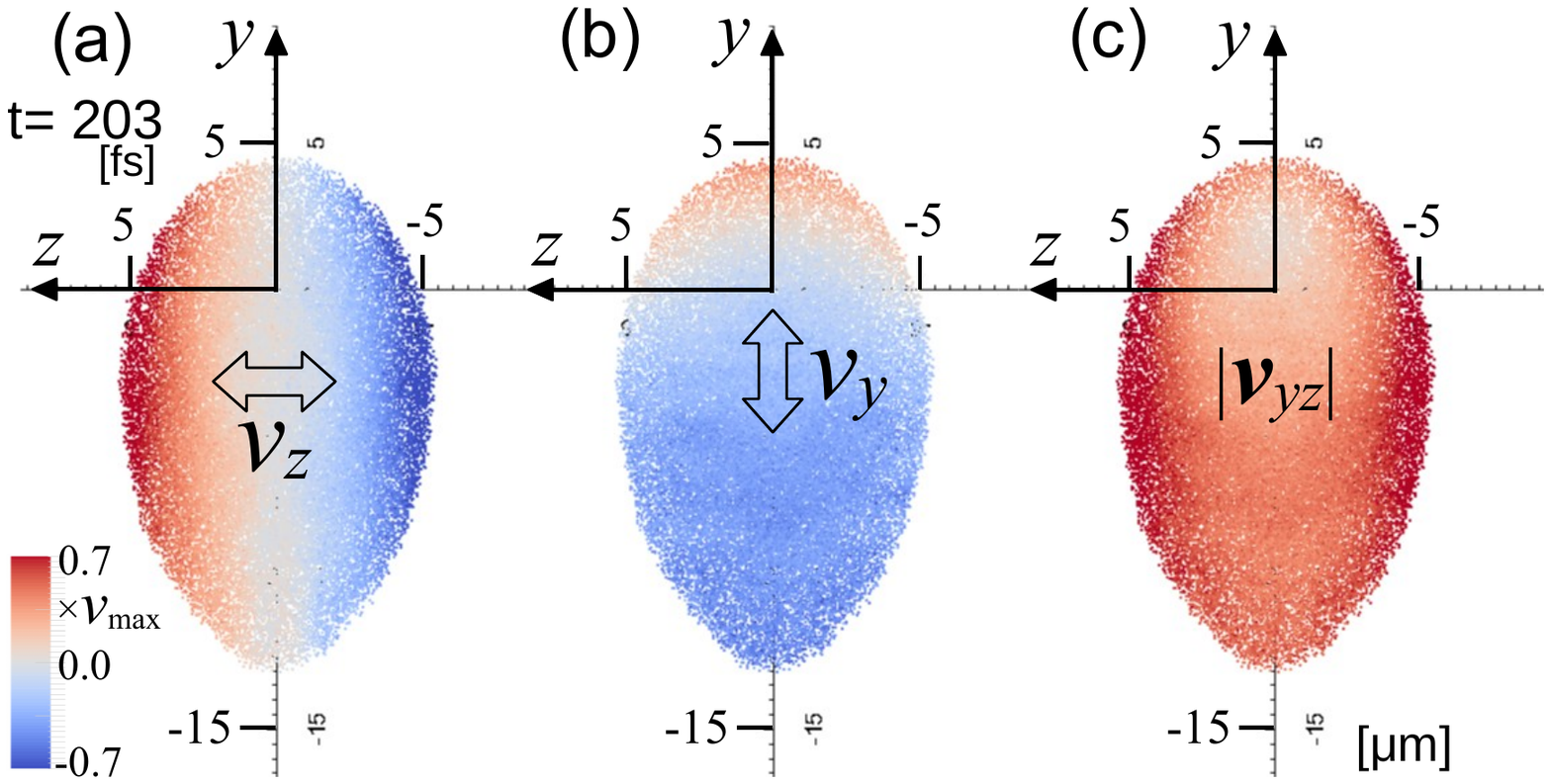}
\caption{
Ion distribution of Fig. \ref{fig:fig_4mev}(b) colored by the velocity of each ion.
Color-coded by horizontal velocity $v_z$ (a), by the vertical velocity $v_y$ (b),
and by their amplitude $|\bm{v}_{yz}|= \sqrt{v_z^2+v_y^2}$ (c) at $t=203$ fs.
Normalized by the maximum $v$ among (a), (b), and (c),
i.e. $\mathrm{max}\{|\bm{v}_{yz}|_i\}$.
In the horizontal direction, it expands by about $1.5$ times the speed
in the vertical direction.
}
\label{fig:fig_v4m}
\end{figure}

The ion distribution in Fig. \ref{fig:fig_4mev}(b), colored by the velocity of each
ion in the $y$ and $z$ directions, is shown in Fig. \ref{fig:fig_v4m}.
The darker color indicates that the ion velocity is faster,
and the contour levels are the same for (a), (b), and (c).
Figure \ref{fig:fig_v4m}(a) is color-coded by the horizontal velocity, $v_z$, with
red representing the velocity in the $+z$ direction and blue in the $-z$ direction.
The horizontal velocity is higher toward the two ends of the bunch and 
is almost zero near its center.
That is, in the horizontal direction, the ion bunch expands with its center as
the immovable point, and the velocity is faster for the ions at both ends of the
horizontal direction.
Figure \ref{fig:fig_v4m}(b) is color-coded by the vertical velocity, $v_y$.
Overall, the color is lighter than that of Fig. \ref{fig:fig_v4m}(a),
i.e. the velocity is lower, and the edges are not noticeably faster.
Moreover, the proportion of the blue part is large,
indicating that the ion bunch has a downward, $-y$ direction, velocity on average.
Therefore, the position that has not moved vertically, i.e. the white area, is
shifted in the $+y$ direction from the center of the ion bunch.
In Fig. \ref{fig:fig_v4m}(c), the ions are colored by their amplitudes $v_z$ and
$v_y$, $|\bm{v}_{yz}|= \sqrt{v_z^2+v_y^2}$.
The darkest color (highest velocity) is at both ends of the horizontal direction,
and the lightest color (lowest velocity) is located at a position slightly off
the $+y$ side from $y=z=0$.
In the horizontal direction,
it is expanded by about $1.5$ times the speed in the vertical direction.
The generated ion bunch has a higher velocity in the horizontal, $z$, direction than
in the vertical, $y$, direction.  That is,
thereafter, the ion bunch expands more significantly in the horizontal direction.
Consequently, we obtain a $4$ MeV/u ion bunch with a large horizontal spread.

Next, we explain why the horizontal spread of the velocity vector is greater than
that of the vertical vector.
In our simulation, we use a $45^{\mathrm{o}}$ obliquely incident laser pulse; thus,
the laser irradiating area on the target is a vertically long ellipse centered
at $y=z=0$ (Fig. \ref{fig:fig_lsr}(b)).
Consequently, the acceleration field $E_x$ is distributed over a shorter distance
in the horizontal, $z$, direction than in the vertical, $y$, direction
(Fig. \ref{fig:fig_lsr}(c)).
Thus, $E_x$ changes to a certain value at a short distance in the $z$ direction,
and changes to that value at a longer distance in the $y$ direction, i.e.,
$|\partial E_x/\partial z| > |\partial E_x/\partial y|$.
The ions are accelerated by $E_x$; thus, the larger the $E_x$ position,
the larger the amount of ion movement in the $x$ direction, $\delta_x$.
Therefore,
$|\partial \delta_x/\partial z| > |\partial \delta_x/\partial y|$,
i.e. the curve of the target surface is steeper in the horizontal, $z$, direction
than in the vertical, $y$, direction in the early stage of acceleration.
This is the reason for the higher velocity in the horizontal direction than
in the vertical direction.
In our simulation, because of the $45^{\mathrm{o}}$ oblique incidence,
the FWHM of the laser intensity in the vertical direction on the target is longer than
in the horizontal direction by a factor of $\sqrt{2}$.
This can also be expressed as the spot size of the laser in the horizontal direction
is smaller than that in the vertical direction.
These results demonstrate that the divergence of the ions in the direction of the
short spot size is large.
This indicates that for two circularly focused laser pulses with different spot sizes,
the case with a shorter spot size has a larger divergence of the generated ions than
the case with a larger spot size.

\begin{figure}[tbp]
\includegraphics[clip,width=8.0cm,bb=32 32 432 205]{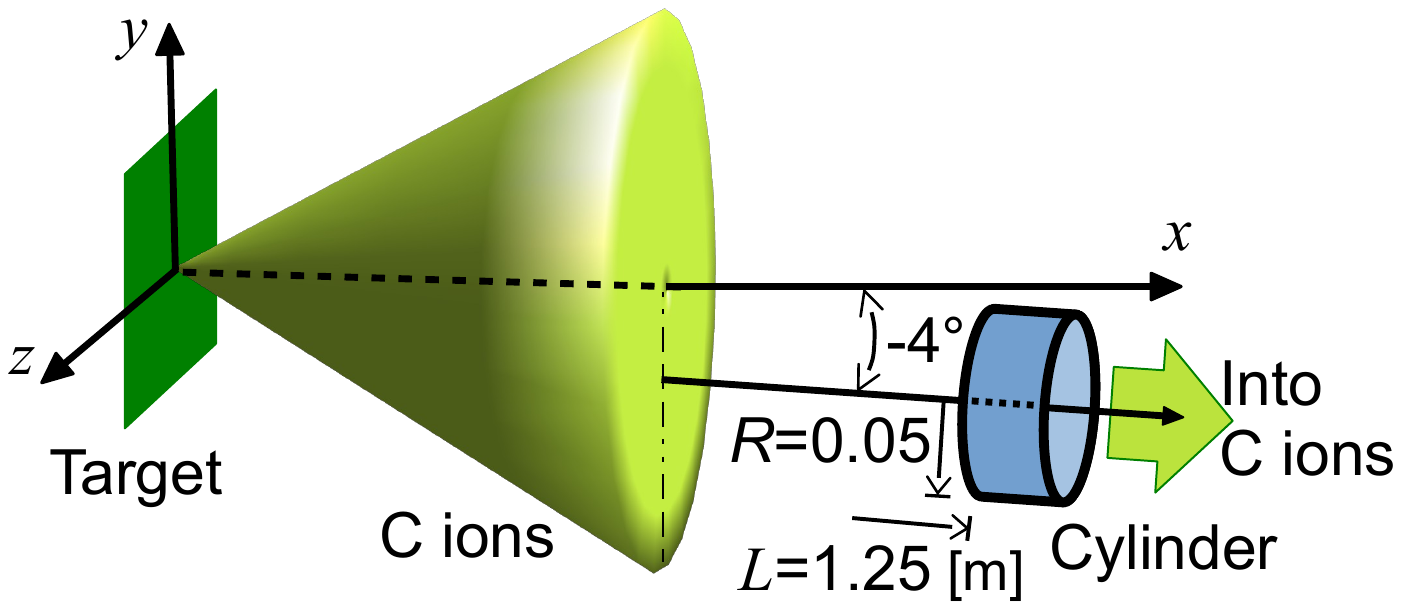}
\caption{
Illustration of position of the cylinder and the target.
The cylinder is placed along a line inclined $4^{\mathrm{o}}$ downward
from the $x$ axis, and the center line of the cylinder is on that line.
}
\label{fig:fig_cyl}
\end{figure}

\begin{figure}[tbp]
\includegraphics[clip,width=8.0cm,bb=37 37 543 360]{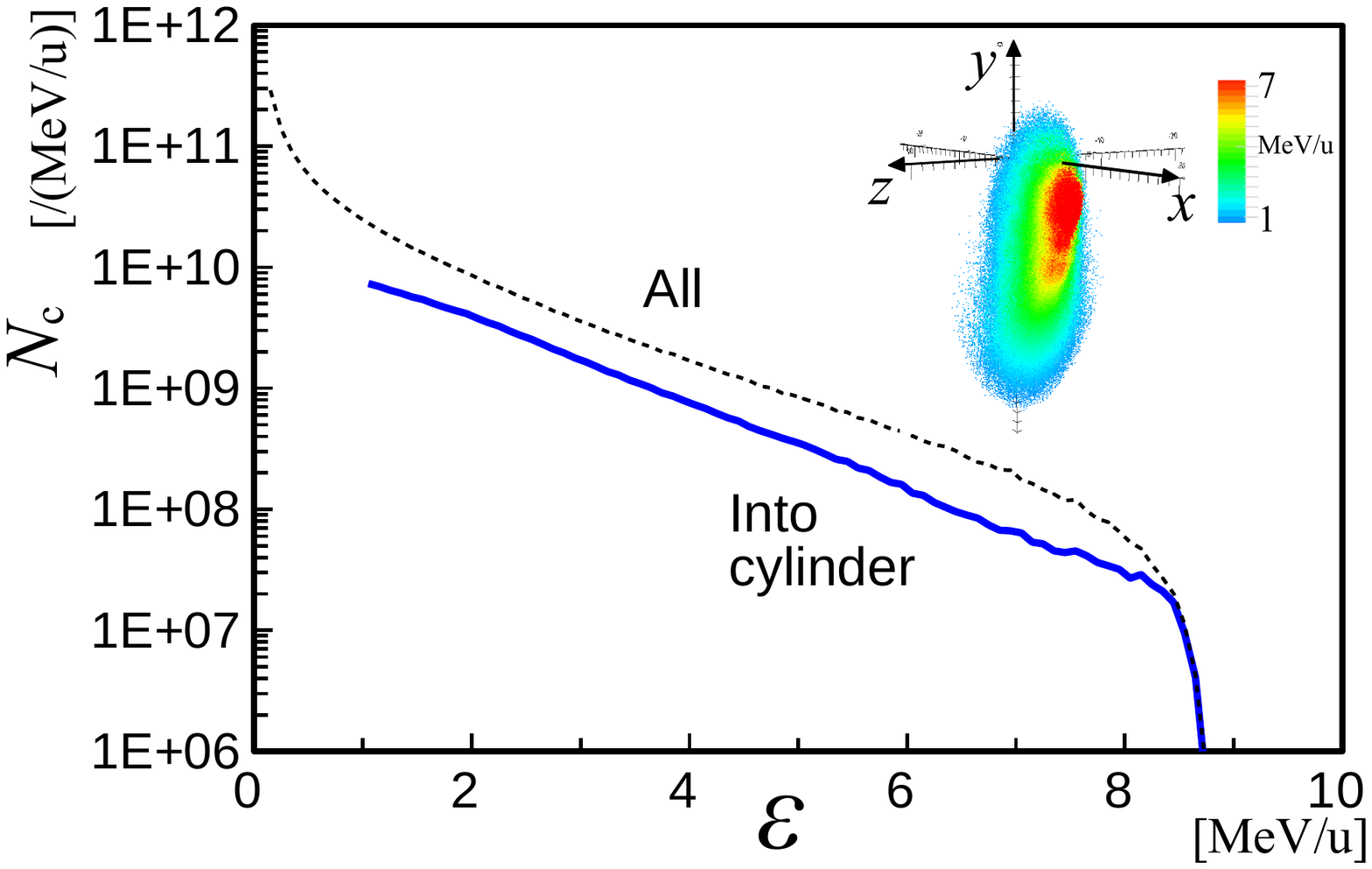}
\caption{
Energy spectrum of the carbon ions coming into the cylinder in the simulation
at $t=203$ fs is marked by a solid line.
In the inset: spatial distribution of those ions.
The dotted line is for all ions accelerated in the $+x$ direction.
}
\label{fig:fig_hole_es}
\end{figure}

Generally, in applications, ions generated by laser acceleration are received
and used in equipment installed in the latter stage.
In this study,
we assume that the entrance of the generated ions in the subsequent device is
a $10$ cm diameter circular hole located $1.25$ m back from the target, i.e.
$5$ msr in solid angle (Fig. \ref{fig:fig_cyl}).
It is assumed that the ions accelerated in the $+x$ direction and within that range
are utilized.
Therefore,
we will examine in detail the characteristics of ions with energy of $4$ MeV/u and
within a solid angle of $5$ msr.
Based on the results of the $\varphi$ distribution in Fig. \ref{fig:fig_theta},
the circular hole that receives the accelerated ions is placed at an inclination
of $-4^{\mathrm{o}}$ from the $x$ axis with its face toward the center of the target
to receive the largest number of ions (Fig. \ref{fig:fig_cyl}).
Figure \ref{fig:fig_hole_es} shows the energy spectrum of carbon ions passing through
the circular hole.
Here, ion passing is defined as the extension of the momentum vector starting from
the position of the ion into the circle at $t=203$ fs.
The result for ions at energies greater than $1$ MeV/u is shown by a solid line.
The spatial distribution of these ions is shown in the inset.
The result for all ions accelerated in the $+x$ direction (i.e. the result in
Fig. \ref{fig:fig_espc}) is also shown by a dotted line for comparison.
The number of ions passing through the circle is about $40\%$ of the all,
except near the maximum energy.
Almost all of the ions near the maximum energy pass through the circle due to the
inclined arrangement of the circle.
For the $4$ MeV/u ($3.8$ -- $4.2$ MeV/u) ions,
$3 \times 10^8$ ions have passed through the circle.
That is, $6 \times 10^7$ ions are obtained per $1$ msr,
which is a sufficiently large number for some applications.
Although at around the maximum energy, $8$ MeV/u, the number of ions with the same
energy range, $7.8$ -- $8.2$ MeV/u, passing it is $2 \times 10^6$ /msr,
which is significantly less than that at around $4$ MeV/u, and is about $1/30$.
This is the reason for our focus on lower energy ions,
at around $4$ MeV/u, instead of near the maximum energy.

\begin{figure}[tbp]
\includegraphics[clip,width=7.0cm,bb=32 40 327 435]{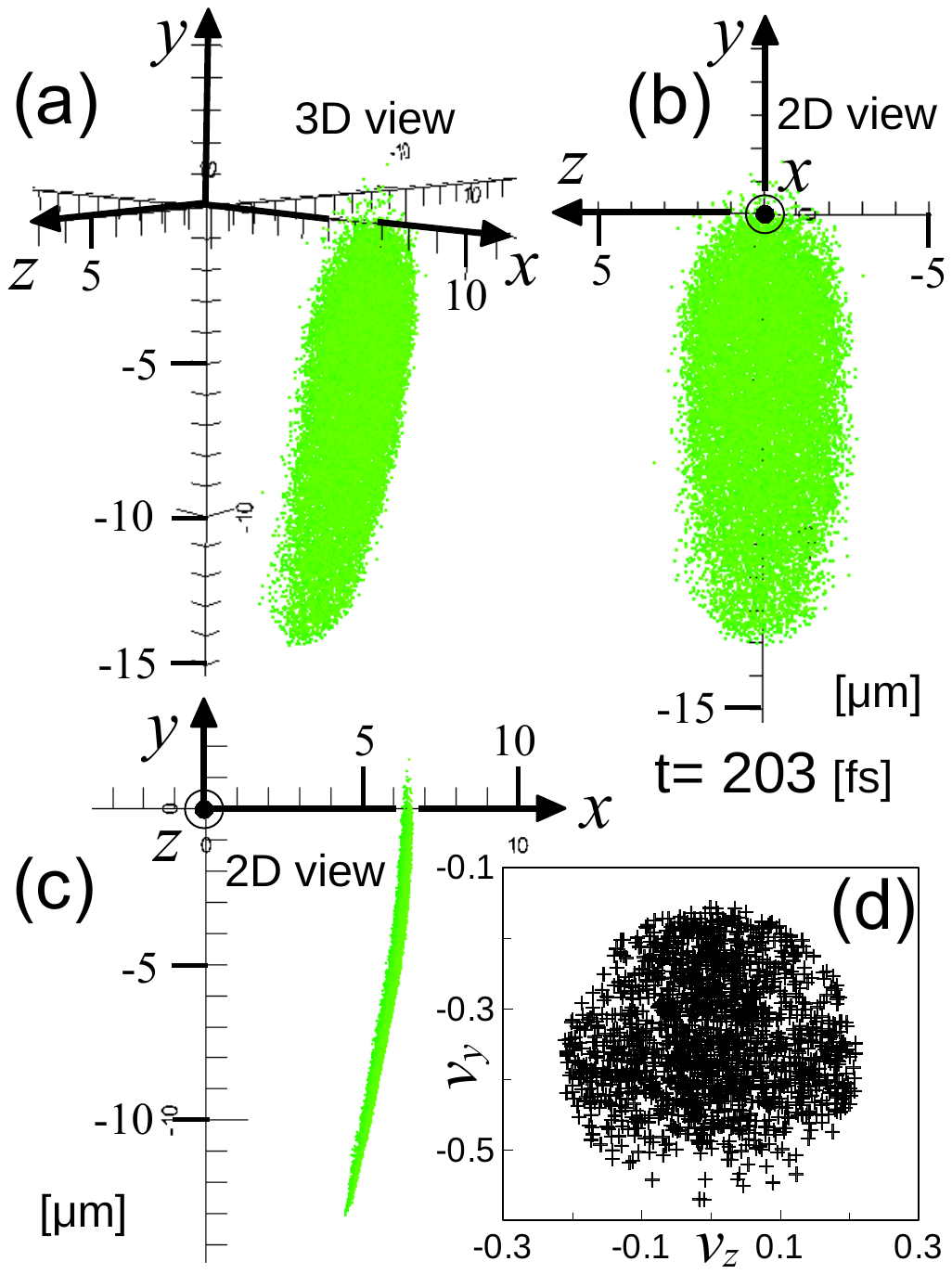}
\caption{
Spatial distribution of the $4$ MeV/u carbon ions at $t=203$ fs that
come into the cylinder.
(a) 3D view.
2D view as viewed along the $x$ axis (b),
and along the $z$ axis (c).
The ions are distributed vertically elongated.
(d) Scatter plot of those carbon ion velocities, $v_z$ versus $v_y$,
normalized by the maximum value of $|v_z|$ among whole $4$ MeV/u ions,
i.e. all ions in Fig. \ref{fig:fig_4mev}.
}
\label{fig:fig_hole_4m}
\end{figure}

Figure \ref{fig:fig_hole_4m}(a,b,c) show the spatial distributions of the $4$ MeV/u
carbon ions at $t=203$ fs that pass through the circle. (a,b,c,d) correspond to
(a,b,c,d) in Fig. \ref{fig:fig_4mev}.
That is, Fig. \ref{fig:fig_hole_4m}(a) shows a 3D view,
Figs. \ref{fig:fig_hole_4m}(b) and (c) are 2D projections as viewed along the $x$ and
$z$ axes, respectively, and
Fig. \ref{fig:fig_hole_4m}(d) is a scatter plot of the ions in $v_z \times v_y$,
normalized by the maximum $v$ among all $4$ MeV/u ions, i.e. all ions
in Fig. \ref{fig:fig_4mev}.
The thickness of the ion bunch is very thin and has a shell-like shape spread along
the $yz-$plane.
The ion bunch is shifted more toward the $-y$ direction compared to that
in Fig. \ref{fig:fig_4mev} because the circular hole is shifted to $-y$.
The distribution of ions is elongated vertically, as shown
in Fig. \ref{fig:fig_hole_4m}(b).
The ion distribution through the circle is extremely vertically long because the
horizontal, $z$ direction, spread of the velocity vector is larger than the vertical,
$y$ direction, spread, as shown in Fig. \ref{fig:fig_v4m}.
That is, the ions around the horizontal edges of the ion bunch shown in
Fig. \ref{fig:fig_v4m} do not pass through the circle because of their high horizontal
velocity, and the area where the horizontal edges are largely cut off is
the ion bunch that passes through the circle.
The scatter plot in Fig. \ref{fig:fig_hole_4m}(d) shows a nearly circular
distribution, as opposed to the long horizontal distribution
in Fig. \ref{fig:fig_4mev}(d).
This is because, for the $10$ cm diameter circular hole placed $1.25$ m back,
this ion bunch is practically almost a point, and the distance from the position of
each ion to the edge of the circular hole is approximately the same for all ions.
Therefore, the ions passing through the circular hole satisfy the condition
$\sqrt{v_y^2+v_z^2} < r/t$,
where $r$ is the radius of the circular hole and
$t$ is the time required by the ions to reach the edge of the circular hole,
i.e. a circular distribution is formed.

\section{CONCLUSIONS} \label{con}

Carbon ion acceleration driven by a laser pulse irradiating a carbon foil target
is investigated with the help of 3D PIC simulations.
In ion acceleration using low-energy laser pulses with solid foil targets,
the acceleration process is Coulomb explosion.
The ion clouds generated by this scheme have a layered structure with
different energies, such as the layers of an onion.
An ion bunch extracted with a narrow energy range becomes a thin shell shape
with a certain diameter; the higher its energy, the shorter its diameter.
The ion energy spectrum monotonically decreases; thus,
the lower the energy of the ion, the greater its number.
In our simulation, the number of ions generated at around half the maximum energy
is approximately $30$ times that near the maximum energy.
For applications that require a large number of ions,
it is advisable to use ions with lower energy than those with the maximum energy.

In the oblique incidence laser, the divergence of the generated ions is smaller
in the laser inclination direction than in the direction perpendicular to it.
That is, the divergence of the generated ions is larger for lasers with
shorter spot sizes than for those with longer spot sizes.
The spot size of the laser has a significant effect on the divergence of
the obtained ions.

The radius of the obtained ion bunch can be expressed in terms of its energy,
the maximum generated ion energy,
and a coefficient representing the laser intensity distribution.
The ion energy spectrum can be expressed by the initial density of the target,
and the thickness which becomes accelerated ions in the target, in addition to them.
From these formulae, we can obtain important indicators for the design of applications.

\section*{ACKNOWLEDGMENTS}
I thank S. V. Bulanov, T. Zh. Esirkepov, M. Kando, T. Kawachi, J. Koga
and K. Kondo for their valuable discussions.
This work was supported by JST-MIRAI R \& D Program Grant Number JPMJ17A1.
The computations were performed with the supercomputer HPE SGI8600 at JAEA Tokai.


\end{document}